\documentclass[fleqn,usenatbib]{mnras}

\usepackage{newtxtext,newtxmath}

\usepackage[T1]{fontenc}

\DeclareRobustCommand{\VAN}[3]{#2}
\let\VANthebibliography\thebibliography
\def\thebibliography{\DeclareRobustCommand{\VAN}[3]{##3}\VANthebibliography}

\usepackage{graphicx}	
\usepackage{amsmath}	
\usepackage{xcolor}
\usepackage[normalem]{ulem}

\usepackage{ifthen}
\newboolean{comments}
\setboolean{comments}{true}

\ifthenelse{\boolean{comments}}
{
\newcommand{\edit}[1]{\textbf{\textcolor{magenta}{ #1}}}
}
{
\newcommand{\edit}[1]{}
}

\usepackage[normalem]{ulem}

\usepackage[acronym]{glossaries}

\newacronym{gw}{GW}{gravitational wave}
\newacronym{lvk}{LVK}{LIGO-Virgo-KAGRA}
\newacronym{snr}{SNR}{signal to noise ratio}

\newacronym{bh}{BH}{black hole}
\newacronym{ns}{NS}{neutron star}
\newacronym{bbh}{BBH}{binary black hole}

\newacronym{nsbh}{NSBH}{neutron star -- black hole}
\newacronym{bhns}{BHNS}{black hole -- neutron star}
\glsdisablehyper

\usepackage{hyperref}
\usepackage{academicons}

\title[Astrophysical Inference for GW230529]{The Impact of Astrophysical  Priors on Parameter Inference for GW230529}

\author[D. Chattopadhyay et al.]{Debatri Chattopadhyay,$^{1}$\thanks{E-mail: ChattopadhyayD@cardiff.ac.uk}
Sama Al-Shammari$^{1}$
Fabio Antonini,$^{1}$
Stephen Fairhurst,$^{1}$
\newauthor
 Benjamin Miles,$^{1}$
and Vivien Raymond$^{1}$
\\
$^{1}$School of Physics and Astronomy, Cardiff University, Cardiff, CF24 3AA, United Kingdom
}

\date{Accepted XXX. Received YYY; in original form ZZZ}

\pubyear{2024}

\begin{document}
\label{firstpage}
\pagerange{\pageref{firstpage}--\pageref{lastpage}}
\maketitle

\begin{abstract}
We investigate the effects of prior selection on the inferred mass and spin parameters of the neutron star-black hole merger GW230529\_181500. Specifically, we explore models motivated by astrophysical considerations, including massive binary and pulsar evolution. We examine mass and spin distributions of neutron stars constrained by radio pulsar observations, alongside black hole spin observations from previous gravitational wave detections. We show that the inferred mass distribution highly depends upon the spin prior. 
Specifically, under the most restrictive, binary stellar evolution models, we obtain narrower distributions of masses with a black hole mass of $4.3^{+0.1}_{-0.1}\,M_{\odot}$and neutron star mass of $1.3^{+0.03}_{-0.03}\,M_{\odot}$ where, somewhat surprisingly, it is the prior on component spins which has the greatest impact on the inferred mass distributions. Re-weighting using neutron star mass and spin priors from observations of radio pulsars, with black hole spins from observations of gravitational waves, yields the black hole and the neutron star masses to be  $3.8^{+0.5}_{-0.6} \,M_\odot$ and $1.4^{+0.2}_{-0.1} \,M_\odot$ respectively. The sequence of compact object formation --- whether the neutron star or the black hole formed first --- cannot be determined at the observed signal-to-noise ratio. However, there is no evidence that the black hole was tidally spun up.
\end{abstract}




\section{Introduction}
\label{sec:intro}

The \gls{gw} event GW230529\_181500 (hereafter abbreviated GW230529) was detected by the LIGO Livingston Observatory during the first part of the fourth observing run of the LIGO-Virgo-KAGRA collaboration \citep{gw230529}. The binary components are a neutron star and another compact object whose mass likely falls within the ``lower mass gap'' between the heaviest neutron stars and the lightest black holes, which was previously thought to exist between $2.5\,M_\odot$ to approximately $5\,M_\odot$ \citep{Bailyn1998, Ozel2010,2011ApJ...741..103F}. Moreover, the event marks the third confident detection of a black hole-neutron star binary through gravitational waves \citep{LIGOScientific:2021qlt}. It followed a probable observation of a pulsar-(mass gap) black hole in the globular cluster NGC 1851 \citep{Barr2024}. 
Theoretical astrophysics of massive binary evolution has speculated about the existence of black hole-neutron star binaries \citep{Oshaughnessy2008, Fryer:2012, Chattopadhyay2021, Broekgaarden2021}. Still, successive detections of such objects spark more debates on their mass-spin distributions to formation channels \citep{Chandra:2024} to possible electromagnetic counterparts \citep{Zhu:2024, Ronchini:2024}.

The detection of GW230529 has significant implications for our understanding of stellar evolution and the end stages of massive stars, 
since it might provide further evidence for compact objects existing within the mass gap \citep{Zevin:2020,Siegel:2023,Zhu:2024,2024arXiv240506819M}. 
In the pre-gravitational wave era, this perceived mass gap led theoreticians to speculate on the supernova mechanisms.
For example, the rapid and delayed supernova models discussed in  \citet{Fryer:2012} present different growth timescales for the instabilities driving the explosion of massive stars. The rapid supernova model can reproduce the mass gap, suggesting that the formation of \glspl{ns} and \glspl{bh} occurs within distinct mass ranges. In contrast, the delayed supernova model, where supernova explosions occur long after the bounce shock, predicts that the mass gap will be populated. The delayed model will therefore imply a smoother transition in the remnant masses of supernovae and allow for the existence of compact objects with masses in the gap \citep{Belczynski2012,Olejak:2022}.

The inherent limitations in recovering the binary physical parameters from the gravitational wave data hinder the accurate measurement of the component masses and spin parameters in a gravitational wave event like GW230529.
While the component masses and spins are the most (astro-)physically interesting quantities, they are not the variables which most directly affect the emitted gravitational wave.  Those are the chirp mass, $\mathcal{M}$, mass ratio $q$, and effective spin, $\chi_{\mathrm{eff}}$ (which we define in Section \ref{sec:methods}).  For a low-mass system like GW230529, the chirp mass is measured with good accuracy, with $\mathcal{M} = 1.94 \pm 0.04 M_{\odot}$, while there is significantly larger uncertainty on the mass ratio and effective spins \citep{gw230529}.  Furthermore, as is well known \citep{Cutler:1994ys, Hannam:2013uu}, there is a degeneracy between the measured mass ratio values and effective spin. Consequently, changing the prior assumptions for spins of the binary component spins can significantly impact the inferred mass distributions and, conversely, changing the mass assumptions can impact the inferred spins.  In this work, we investigate the impact of using astrophysically motivated distributions for the masses and spins of \glspl{bh} and \glspl{ns} on the inferences about the progenitor properties of GW230529.

The remainder of the paper is laid out as follows: in Section \ref{sec:methods} we briefly summarize the methods used to incorporate astrophysical priors, in \ref{sec:models} we introduce the astrophysical observations and models used, in \ref{sec:results} we present the results and we conclude in \ref{sec:discussion} with a discussion and possible future directions.

\section{Incorporating Astrophysical Prior Beliefs}
\label{sec:methods}
The frequency evolution of a \gls{gw} emitted by an inspiralling compact binary is determined, at leading order, by the chirp mass, $\mathcal{M}$, of the system, while the mass ratio $q$, and effective spin, $\chi_{\mathrm{eff}}$, affect the signal at sub-leading orders \citep{Blanchet:2006zz}.%
\footnote{The precessing spin, $\chi_{\mathrm{p}}$, also impacts the observed waveform.  However, as there is no evidence for precession in this observation and the in-plane spins are essentially unconstrained (see Figure 13 of \cite{gw230529}) we do not consider the precessing spin here}
These quantities are related to the component masses ($m_{1, 2}$) and (spins $\mathbf{S}_{1, 2}$):
\begin{align}
    \mathcal{M} &= \frac{(m_{1} m_{2})^{3/5}}{(m_{1} + m_{2})^{{1/5}}} \\
    q &= \frac{m_2}{m_1} \\
    \chi_{\mathrm{eff}} &= \frac{(m_{1} \boldsymbol{\chi}_{1} + m_{2} \boldsymbol{\chi}_{2}) \cdot \mathbf{\hat{L}}}
    {m_{1} + m_{2}} \label{eq:chi_eff}
\end{align}
where $\mathbf{\hat{L}}$ is the unit vector in the direction of the orbital angular momentum, and the spin, $\boldsymbol{\chi}_{i}$, is defined as $\boldsymbol{\chi}_{i} = \frac{\mathbf{S}_{i}}{m_{i}^{2}}$, and the normalization is chosen such that a maximally spinning \gls{bh} has $|\boldsymbol{\chi}| = 1$.   Consequently, the inferred values of the component masses and spins from a \gls{gw} observation correlate.  Furthermore, the choice of priors can impact the inferred properties for a low \gls{snr} signal such as GW230529.  Indeed, due to correlations between the inferred mass ratio and effective spin, choices of spin priors can impact the inferred masses and vice versa.

The gravitational-wave estimation of these parameters for GW230529, as presented in \cite{gw230529}, was performed with minimally informed priors on the mass and spin distributions of the components of the binary, in addition to the location and orientation parameters \citep{Veitch:2014wba}.  Specifically, the analysis used mass priors that are flat in the redshifted component masses, within a prescribed range of masses and mass ratios, and uniform in spin magnitude and orientation (see Appendix D of \cite{gw230529} for details).  The output of the original analysis is a set of samples from the posterior probability distribution of the parameters \citep{ligo_scientific_collaboration_2024_10845779}%
\footnote{Specifically, we make use of the \texttt{Combined\_PHM\_highSpin} results for the majority our our analyses.  When re-weighting using the astrophysical models described in Section \ref{sec:modelA}, which restrict component spins to be close to zero, we use \texttt{IMRPhenomPv2\_NRTidalv2\_lowSpin} to ensure that we have sufficient samples following re-weighting}
.
We then apply weights to these samples in the ratio of our desired \textit{astrophysical} before the \textit{original}, uniform prior.  Given an astrophysically motivated prior $\pi_{A}$, the posterior distribution for the parameters $\boldsymbol{\theta}$ describing the progenitor of GW230529 is given by
\begin{equation}\label{eq:posterior}
    p_{A}(\boldsymbol{\theta} | d) 
    = \frac{\pi_{A}(\boldsymbol{\theta}) p(d | \boldsymbol{\theta})}{p_{A}(d)}
\end{equation}
where $p(d | \boldsymbol{\theta})$ is the likelihood of the data given the parameters $\boldsymbol{\theta}$.  The denominator, $p_{A}(d)$ is the evidence which serves as an overall normalization of the distribution.  

While it is possible to calculate this posterior directly from the data, it is more straightforward to \textit{re-weight} the existing results obtained with the original, uninformed prior $\pi_{0}$, see \cite{2019PhRvD.100l3017P} for a demonstration.  In particular, we can simply re-weight the distribution to obtain the posterior associated with the astrophysical prior:
\begin{equation}\label{eq:astro_post}
    p_{A}(\boldsymbol{\theta} | d) 
    = \frac{\pi_{A}(\boldsymbol{\theta})}{\pi_{0}(\boldsymbol{\theta})} 
    p_{0}(\boldsymbol{\theta} | d) \, .
\end{equation}
The posterior distribution is provided as a set of discrete samples $\theta_{i}$ whose density in parameter space follows the posterior distribution, $p_{0}(\boldsymbol{\theta} | d)$.  Thus, to obtain samples associated with the astrophysical prior, we calculate a weighting factor for each sample, 
\begin{equation}
    w_{i} = \frac{\pi_{A}(\boldsymbol{\theta}_{i})}{\pi_{0}(\boldsymbol{\theta}_{i})} \ .
\end{equation}
This set of weighted samples provides updated parameter estimates under astrophysically motivated prior assumptions.  To obtain a discrete set of equally weighted samples, we perform importance sampling on the weighted samples.  This re-weighting is performed using importance sampling \citep{goertzel1950quota,Robert_Casella,Liu}. To do this, we calculate the maximum weight $w_{\mathrm{max}}$ across the samples and then keep each of the samples $\boldsymbol{\theta}_{i}$ with a probability
\begin{equation}
    p_i = \frac{w_{i}}{w_{\mathrm{max}}} \, .
\end{equation}

When performing prior re-weighting, however, it is important that the proposed prior provides support in the same part of the parameter space as the target prior, otherwise the re-weighting will be highly inefficient.  Since the original priors are near-uniform in the parameters of interest and also cover a broad parameter range, the astrophysical priors generally comprise a subset of the original ranges.  However, in cases where the astrophysical priors are sharply peaked, the re-weighting procedure summarized above can lead to a low number of samples in the astrophysical posterior. In our analyses we ensured a minimum of one thousand effective samples \citep{Kish, Elvira}. For the results presented in Section \ref{sec:results}, we use those samples to create one-dimensional Kernel Density Estimators of the inferred masses and spins. As 90\% credible intervals are the relevant numbers for the scientific conclusions of this work, we checked via bootstrapping tests to what precision the one-dimensional credible intervals could be quoted, and used that precision.

The astrophysical models introduced in the next section provide informed distributions for a subset of the parameters of the binary.  We have no reason to use an informative prior for the sky location or orientation of the binary, so we do not re-weight these parameters in any of our studies.  
We are, however, interested in re-weighting (a subset of) the mass and spin parameters.  Since the original priors for the masses, spin magnitudes and orientations are independent, we are free to perform the re-weighting procedure described above separately for each parameter, provided that our astrophysically-motivated priors are also independent in these parameters.  In all cases, only a subset of the parameters are re-weighted while the others retain the original, uninformative priors.  Then, the re-weighting factorizes as
\begin{equation}
    w_{i} =  \frac{\pi_{A}(\theta_{1}) \pi_{A}(\theta_{2}) \ldots \pi_{A}(\theta_{n}) } {\pi_{0}(\theta_{1}) \pi_{0}(\theta_{2}) \ldots \pi_{0}(\theta_{n})} \, .
\end{equation}
When re-weighting the masses, we introduce new prior distributions for either $m_{1}$, $m_{2}$ or both.  When re-weighting the spins, we use astrophysically motivated distributions for one or both of the spin magnitudes $|\boldsymbol{\chi}_{i}|$, or both the magnitudes and orientation through their impact upon the z-component of the spin
\begin{equation}
    \chi_{z} = \boldsymbol{\chi} \cdot \mathbf{\hat{L}} \, .
\end{equation}

\section{Models}
\label{sec:models}

\begin{table*}
\begin{tabular}{llllllll}
Model                & Description          & M$_\mathrm{BH}$        & M$_\mathrm{NS}$                & $\chi_\mathrm{BH}$          & $\chi_\mathrm{NS}$     & tilt BH & tilt NS 
\\
\hline
LVK  & uninformed priors$^\ast$ & flat in det. frame$^\ast$  & flat in det. frame$^\ast$ & uniform [0-1]  & uniform [0-1] & uniform & uniform          \\
FDT500 & astro. model$^\dagger$& fit to model  & fit to model& $\approx 0$  & $< 10^{-2}$ & aligned & aligned \\
FDT500\_Q & astro. model with BH tidal spin-up$^\ddagger$ &fit to model   &  fit to model & fit to model  &  $< 10^{-2}$ & aligned & aligned\\
Pulsar & pulsar mass + GC millisecond pulsar spin$^\mathparagraph$ &uniform  & fit to pulsars & uniform [0-1] & fit to pulsars & uniform & uniform \\
BBH & BBH inferred mass, spin from GWTC-3$^\mathsection$ &uniform  & uniform  & BBH $\chi$  & uniform [0-1] & BBH tilt & uniform\\
Pulsar + BBH & pulsar mass, spin + GW BBH mass, spin &uniform  & fit to pulsars & BBH $\chi$  & fit to pulsars & BBH tilt & uniform \\
\hline
\end{tabular}
\caption{The models we use for re-weighting, as described in detail in Section~\ref{sec:models}. \\ 
$\ast$ identical to \protect\cite{gw230529} ``primary combined analysis"\protect\\ 
$\dagger$ from \protect\cite{Chattopadhyay2021}, $Z=0.02$, common envelope optimistic, $\alpha=1$, rapid supernovae, $\tau_\mathrm{d}=500$Myrs, $\Delta$M$_\mathrm{d}=0.2\, M_\odot$\protect\\
$\ddagger$ as previous, with \gls{bh} tidal spin up \protect\cite{Qin:2018nuz} \protect\\ 
$\mathparagraph$ observed radio pulsar mass from \protect\cite{Rocha2023}, globular cluster (GC) millisecond pulsar spins from ATNF catalogue \protect\\
$\mathsection$ from \protect\cite{KAGRA:2021duu} \gls{bbh} mass, spin distribution\protect\\
}
\label{tab:models}
\end{table*}

The astrophysically motivated distributions of masses and spins that we use later to inform our prior expectations for GW230529 are split into four distinct classes:

\begin{enumerate}
\item Binary population synthesis models.  These models are guided by existing theoretical and observational constraints on binary stellar evolution to produce merging binaries comprising a \gls{ns} and a \gls{bh}.

\item Models informed by radio pulsar observations.  We make use of the masses and spins of \glspl{ns} observed as millisecond pulsars to restrict the prior distributions of the \gls{ns} mass and spin. The \gls{bh} mass and spin are left agnostic. 

\item Models informed by gravitational-wave observations.  The number of neutron star-black hole mergers observed before GW230529 was small, therefore the inferred population properties are only weakly constrained.  We instead consider the inferred \gls{bh} spin distribution from the observed \textit{\gls{bbh} population} \citep{KAGRA:2021duu} as a prior for the \gls{bh} spin in GW230529. The \gls{ns} mass and spin are left agnostic.

\item Models informed by observations of \glspl{ns} and \glspl{bh}.  Finally, as the second and third classes of models independently constrain the \gls{ns} parameters and \gls{bh} spins, we apply both constraints concurrently.

\end{enumerate}

\begin{figure*}
\includegraphics[width=0.8\linewidth]{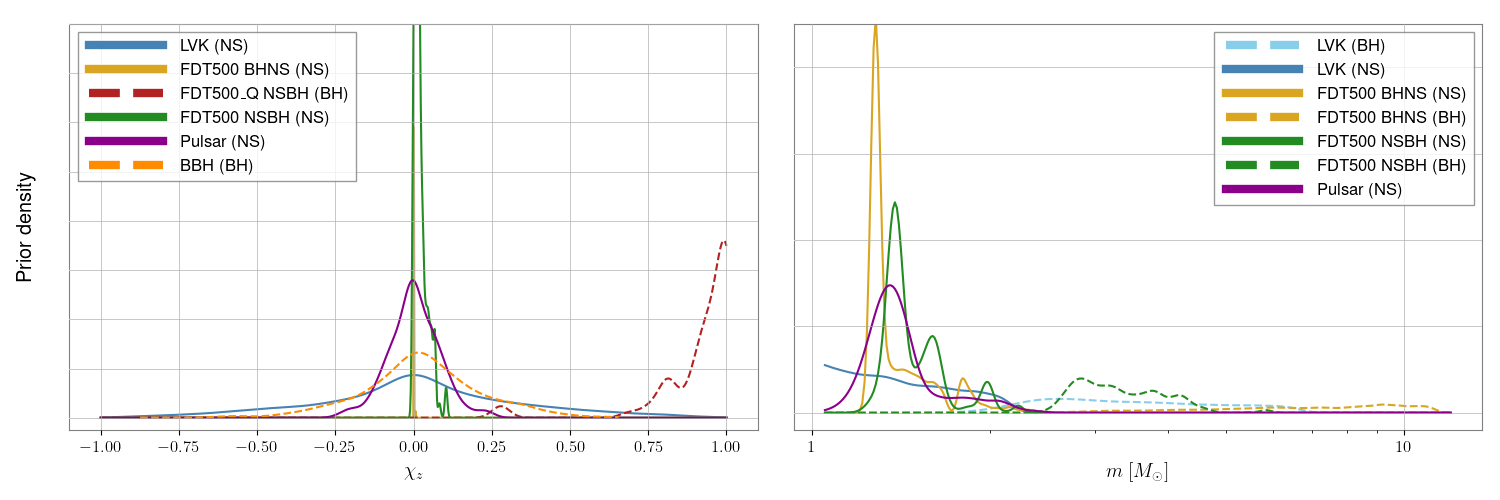}
\caption{The prior distributions for the models outlined in Section \ref{sec:models}: the astrophysical models~\ref{sec:modelA}, the pulsar population models~\ref{sec:modelB} and the GW population models~\ref{sec:modelC}, as well as the priors used by the LVK analysis. The left-hand plot shows the distributions of the spin projected along the orbital angular momentum $\chi_z$; note that the astrophysical FDT500 \gls{bhns} models predict very small spins, while the astrophysical FDT500\_Q \gls{nsbh}  models predict spins largely incompatible with GW230529. We used the FDT500 models as representative of our astrophysical re-weighting. The pulsar population distribution assumed uniform spin orientation. The right plot shows the mass distributions; note that we are not using the mass distributions from the \gls{bbh} population model in the re-weighting.
}\label{fig:prior}
\end{figure*}

The different models are summarized in Table\,~\ref{tab:models} and the relevant quantities are plotted in Figure\,~\ref{fig:prior}.

\subsection{Models motivated by binary astrophysics}
\label{sec:modelA}

Binary stellar evolution models can be used to predict the mass and spin distributions of \glspl{bh} and \glspl{ns} in merging binaries.  For this paper, we consider a set of astrophysical models generated using the population synthesis code $\tt{COMPAS}$ \citep{COMPAS:2021methodsPaper,SSE_Hurley:2000pk,BSE:2002,Fryer:2012,Stevenson2019}. Our base is the Fiducial model, described in 
 \cite{Chattopadhyay2021} \citep[see also][]{Broekgaarden2021}, expanded to explore variations in metallicity, binary evolution and pulsar evolution assumptions. The details of the prescriptions used to model the pulsar spin-down due to magnetic breaking are outlined in \citealt{Chattopadhyay:2019xye} (Section\,2) and \citealt{Chattopadhyay2021} (Section\,2.3). 

In most situations ($>$90\%), it is expected that the more massive compact object, namely the \gls{bh}, will form first.  However, for low-mass binaries with similar zero-age main sequence masses, the slightly more massive initial star can transfer mass to its companion, causing a mass-ratio reversal. The primary star then can become a \gls{ns} first, followed by the secondary forming a light \gls{bh} \citep[for solar metallicity, median around 3.8\,M$_\odot$, as shown in Table\,3 of][although extended tails]{Chattopadhyay2021}. 
We consider a set of models consisting of the two sub-populations \gls{bhns} or \gls{nsbh} to reflect on the order of formation of the compact objects --- whether the black hole is born first (\gls{bhns}) or the neutron star is born first (\gls{nsbh}). 

Our Fiducial model has metallicity $Z=0.02$;  the common envelope parameter that sets the fraction of orbital energy capable of unbinding the envelope is $\alpha=1$ \citep{XuLi:2010,Ivanova2013}; we adopt the `optimistic' scenario for common envelope evolution 
where a Hertzsprung gap (HG) donor can survive and eject the envelope
\citep[][]{Belczynski:2007ApJ}; we  consider the `delayed' supernova prescription by \cite{Fryer:2012};  we set the 
magnetic field decay time-scale affecting both non-recycled and recycled pulsars, $\tau_d=$1000\,Myrs  \citep[see Equation\,5 of][]{Chattopadhyay:2019xye};
and we take the mass-decay time-scale $\Delta M_\mathrm{d}=0.2\,M_\odot$,  which only affects recycled pulsars  \citep[Equation\,12 of][]{Chattopadhyay:2019xye}.

In the other six models, we change one assumption from the Fiducial model. These changes are --- (i) metallicity $Z=0.005$,  (ii) common envelope $\alpha=3.0$, (iii) `pessimistic' common envelope prescription \citep{Belczynski:2007ApJ}, (iv) `rapid' supernova modelling \citep{Fryer:2012}, (v) $\tau_d=500$\,Myrs,  ensuring a slower spin down of both recycled and non-recycled pulsars and vi) $\Delta M_\mathrm{d}=0.02$\,$M_\odot$.
After testing all seven models (Fiducial, and the six variations), we conclude that the effect on the inferred mass of the compact objects is negligibly small. Model v) with $\tau_d=500$\,Myrs shows the widest \gls{ns} spin priors of all models and hence was chosen to represent the astrophysical model set, with the name ``FDT500" (identical to the original paper \citealt{Chattopadhyay2021}).

While the \gls{ns} spin is computed in detail with spinning down (and up, through mass transfer) of pulsars, the \gls{bh} spins are varied only in the \gls{nsbh} sub-population under the assumption of tidal spin-up of the BH-progenitor by the first-born compact object (\gls{ns}, in our case). 
Due to efficient angular momentum transport from core to envelope in He-star progenitor BHs, BHs are expected usually to be born non-spinning, however, in the case of \glspl{nsbh} where the \gls{bh} is born second, tidal effects from the first-born compact object (NS in this case) can potentially spin the second-born \gls{bh} up at birth \citep{Qin:2018nuz,Bavera:2019}. Lower mass, high spin BHs are also expected to be more efficient at generating electromagnetic counterparts at mergers with \glspl{ns} \citep{Barbieri2020}. Therefore, we also consider a population of \glspl{bh} which form with significant spin, calling it ``FDT500\_Q".  The spinning up of the \gls{bh} is a function of the binary orbital period right before the second supernova and the metallicity of the binary (which determines the masses of the compact objects). The fits are derived from \cite{Qin:2018nuz} models and illustrated in equations \,2 and 3 of \cite{Chattopadhyay2022followup}. 

In figure \ref{fig:prior}, we show the inferred distribution of the z-component of the spin, $\chi_{z}$, for the \gls{ns} and \gls{bh} in the models discussed above. All spins are oriented with the orbital angular momentum, assuming efficient tidal alignments. However, the \gls{bh} spin-up model FDT500\_Q show extremely high spins that are completely unsupported by the data and hence discarded. The non-spinning \gls{bh} model of \glspl{nsbh} is considered for prior choices, \glspl{bhns} always have non-spinning \glspl{bh}.

\subsection{Models motivated by radio pulsar observations in star clusters}
\label{sec:modelB}

As an alternative approach to modelling isolated binary evolution, here we utilize the observed population of Galactic globular cluster radio millisecond pulsar%
\footnote{Over a third of all observed millisecond pulsars appear in Galactic globular clusters, which collectively account for less than 0.05\% of the total number of stars in the Milky Way. These millisecond pulsars, have spins up to $\chi \simeq$0.4, several times or even order(s) of magnitude lower than normal pulsars. Although \gls{ns}-\gls{bh} mergers in star clusters are rare \citep{Ye2020}, and \gls{ns}s spin down significantly over time before the merger,
these high-spin pulsars suggest an alternative formation channel that merits exploration.}
spin distribution from the Australia Telescope National Facility (ATNF) pulsar catalogue\footnote{\url{https://www.atnf.csiro.au/research/pulsar/psrcat/}} \citep{Manchester:2004bp}. From the catalogue, we obtain the pulsar's spin $P$, and get 
\begin{equation}
    \chi_\mathrm{NS}=\frac{2\pi cI}{GPM^2},
\end{equation}
where, $c$ is the velocity of light, $G$ is the universal constant of gravitation, $M$ is the mass of the pulsar (assumed to be 1.4\,$M_\odot$ here) and $I$ is the moment of inertia of the pulsar given by 
\begin{equation}
    I = 0.237 M_\mathrm{NS} R_\mathrm{NS}^2 
    \bigg[1 + 4.2 \frac{M_\mathrm{NS}}{M_\odot} \frac{\mathrm{km}}{R_\mathrm{NS}} 
    + 90\Big(\frac{M_\mathrm{NS}}{M_\odot} \frac{\mathrm{km}}{R_\mathrm{NS}} \Big)^4 
    \bigg] ,
    \label{eq:moment_of_inertia}
\end{equation}
\citep[computed from Equation 12 of][with mass $M_\mathrm{NS}$ =1.4\,$M_\odot$ and radius $R_\mathrm{NS}=12$\,Km]{Lattimer2005}.

Although most pulsars are expected to spin down significantly due to the loss of rotational energy, with decaying magnetic field at the time of merger \citep[Figure 17, 21 of][]{Chattopadhyay2021}, we take the observed \gls{ns} spin distribution as an upper limit. The maximum observed spin $\chi_\mathrm{NS}$ is 0.4, with the primary peak at 0.13 and another at about 0.02.   
The mass distribution for NSs, while dependent on the \gls{ns} equation of state and the assumed Tolman-Oppenheimer-Volkoff limit for theoretical studies, can also be estimated from pulsar observations (with the inclusion of radio selection effects). Even with the limited dataset due to the difficulty in measuring \gls{ns} mass, there have been multiple studies to determine the shape and range of the mass distribution of \glspl{ns} \citep{Antoniadis2016,Alsing2018,Rocha2023}. For this analysis, we assume the \gls{ns} mass distribution to be a double Gaussian with a maximum mass of 2.56\,$M_\odot$, with the bimodal mean peaks at 1.351\,$M_\odot$ and 1.816\,$M_\odot$ as found and described in details in \cite{Rocha2023} (see Table\,3). The mass and spin priors for \glspl{bh} remain uninformed and the \glspl{ns}pins are randomly oriented.

\subsection{Models motivated by GW observation of black hole spins}
\label{sec:modelC}

To date, only a small number of \gls{nsbh} or \gls{bhns} binary mergers have been observed \citep{KAGRA:2021duu}.  This makes it difficult to use the observed properties of these binaries to draw strong inferences about their population properties to be used as prior beliefs when interpreting GW230529.  However, during the first three \gls{lvk} observing runs, close to a hundred \gls{bbh} have been observed and the detailed properties of the \gls{bh} population have been inferred \cite{KAGRA:2021duu}.  The inferred mass of the more massive component of GW230529 lies outside of the observed \gls{bh} mass distribution and therefore the existing population mass distribution cannot be used.  However, we can use the inferred spin distributions inferred from the \gls{bbh} population as a proxy for the \gls{bh} spin distribution in \gls{nsbh}/\gls{bhns} binaries.
The inferred spin distribution of \gls{bh} from observations through GWTC-3 is shown in figure 15 of \cite{KAGRA:2021duu}.  The \gls{bh} spin is modelled through the amplitude $\chi$ and orientation $\cos{\theta}$ of the spin relative to the orbital angular momentum.  

The inferred population is given as a set of model distributions, from which an overall average distribution with uncertainty is derived.  For the analysis presented here, we obtain the set of distributions repeat the re-weighting procedure for each distribution and then average over these draws.  The mass distribution of both components and the spin of the \gls{ns} remain unconstrained.

\subsection{Model jointly motivated by observations of radio pulsars and GW binary black holes}
\label{sec:modelD}

As a final model, we combine the astrophysical observations of pulsars from Section \ref{sec:modelB} with those of \glspl{bh} from Section \ref{sec:modelC} to restrict the properties of both components of the progenitor of GW230529.  
As a caveat, we caution that these are two very distinct astrophysical populations. Nevertheless, we consider this approach, noting that it can still provide valuable insights into how the choice of priors influences the parameter recovery. 
We restrict the mass and spin magnitude of the \gls{ns} from pulsar observations, leaving the spin orientation unconstrained, and
we consider the spin magnitude and orientation of the \gls{bh} from \gls{gw} observations, leaving the \gls{bh} mass unconstrained.

\section{Results}
\label{sec:results}

\begin{figure}
\includegraphics[width=1.\linewidth]{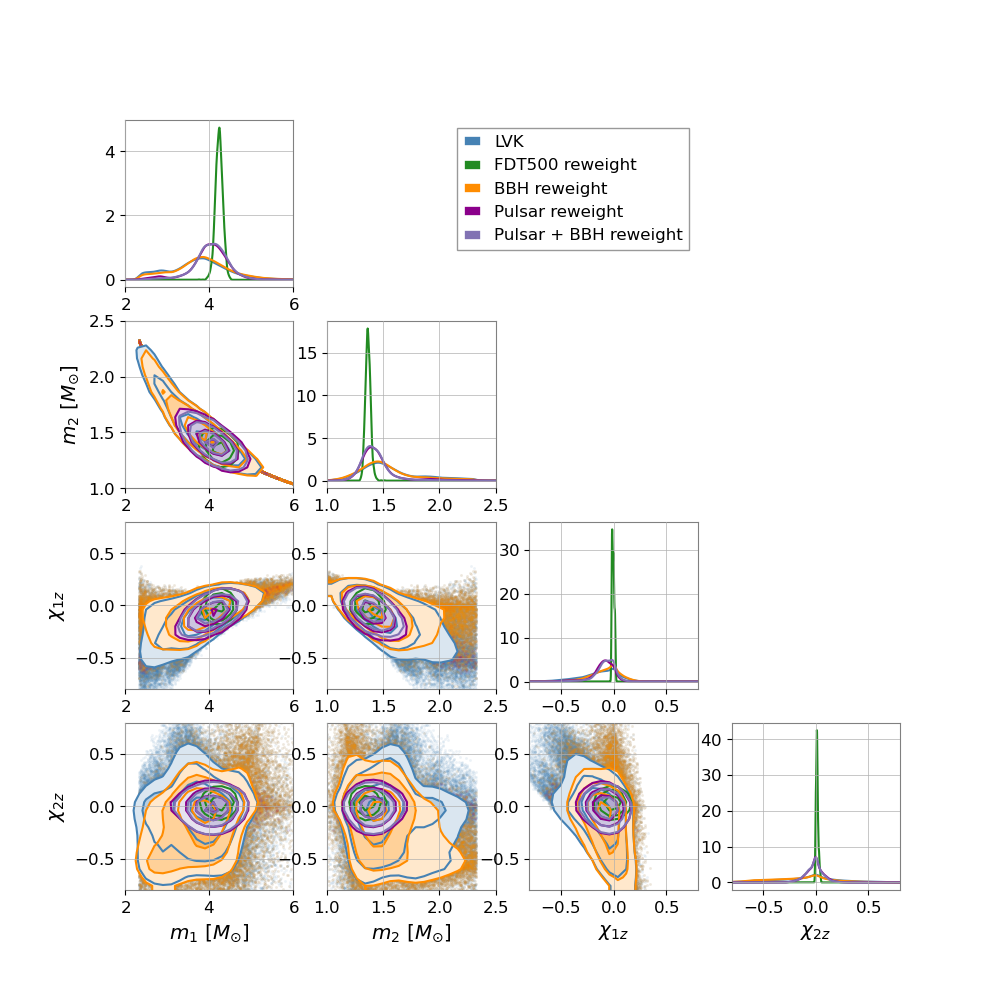}
\caption{The inferred masses and spin projected along the orbital angular momentum $\chi_z$ of the components of GWFch230529, where the subscripts $1$ and $2$, respectively, label the heavier and lighter component of the binary, under five different choices of prior. The \gls{lvk} prior is an uninformative prior which is flat in redshifted masses and uniform in spin magnitude/orientation; the astrophysical FDT500 prior model is discussed in Section \ref{sec:modelA}; the \gls{bbh} prior uses the $\chi_{1z}$ distribution from the observed \gls{bbh} population; the pulsar prior uses masses and spin magnitudes from observed pulsars and the pulsar $+$ \gls{bbh} uses both pulsar and \gls{bbh} observations.  Further details of the models are provided in Section \ref{sec:models} and summarized in Table \ref{tab:models}.  We note that the $\chi_{1z}$ prior for the astrophysical models is so sharply peaked close to 0 that we have included samples within $< 10^{-2}$ to ensure sufficient prior support for re-weighting.
}\label{fig:corner_compare}
\end{figure}

Figure \ref{fig:corner_compare} shows the inferred mass and spin distributions for the primary and secondary components of GW230529 when we impose astrophysically motivated mass and spin priors, shown in Figure \ref{fig:prior}.  In all cases, the imposition of astrophysical prior distributions restricts the inferred range of masses and spins, with the biggest impact coming from the use of astrophysical models and the smallest impact from applying the \gls{bh} spin distribution observed in \gls{bbh}.  For all prior choices, the component masses are more sharply peaked, with reduced support for a close to equal mass, $(2.5, 2.0) M_{\odot}$, system and preference for a \gls{nsbh} with a \gls{bh} mass around $4 M_{\odot}$ and a \gls{ns} mass around $1.4 M_{\odot}$.  Furthermore, we find that the spin distributions narrow, with a preference for low spin magnitudes and reduced support for anti-aligned spins. Interestingly, the results obtained for various astrophysical models are consistent despite these models being used to restrict different subsets of parameters.  

The inferred \gls{ns} and \gls{bh} masses based on the astrophysical model are sharply peaked at $1.3^{+0.03}_{-0.03}\,M_{\odot}$ and $4.3^{+0.1}_{-0.1}\,M_{\odot}$ respectively.  This renders the secondary clearly a \gls{ns} with a mass very compatible with the galactic population and the primary a likely \gls{bh} in the mass gap.  While the object is still in the putative ``lower mass gap", it is towards the upper edge, the supernova explosion does not have to be extremely rapid (and hence energetic) to achieve such mass. Interestingly, the restriction on the binary masses does not arise due to the astrophysical mass distribution, but rather the spin distribution.  As can be seen in Figure \ref{fig:prior}, the \gls{nsbh} mass distribution for both \gls{ns} and \gls{bh} is rather broad, with support from $1.2-2.5\,M_{\odot}$ for \gls{ns} and $2.5 - 6\,M_{\odot}$ for \gls{bh}. The restriction to (close to) zero spins for both components constrains the mass ratio significantly and, consequently, the component masses. To re-iterate, the \glspl{bh} of the BHNSs, are always expected to have zero spin, and the \gls{ns}, being non-recycled, also effectively zero spins. For \glspl{nsbh}, pulsar recycling extends the NS spin distribution to larger values, although by the time of merger, it still spins down to $\lesssim 0.1-0.2$. The \glspl{bh} of the \glspl{nsbh}, have uncertainty associated with their spins --- on one hand we model the optimistic efficient tidal spin-up model which gives spins $\chi_{z} \gtrsim 0.5$ (the FDT500\_Q model in Fig.\ref{fig:prior}), on the other, we assume them to be non-spinning (FDT500). While the tidal spin-up model can be rejected based upon the observed spins, under non-efficient tides we cannot comment definitively on formation order, i.e. \gls{nsbh} vs \gls{bhns}, in this scenario. While \glspl{nsbh} certainly prefers a more symmetric mass ratio and lower \gls{bh} masses, the distribution of both sub-systems is sufficiently broad in mass and peaked near zero spins to be fully consistent with the observation of GW230529.

The \gls{bbh} and pulsar models provide less stringent restrictions on the masses and spins of GW230529.  Indeed, the \gls{bbh} spin prior only restricts the spin of the more massive component and has minimal impact on the masses.  The pulsar observations, and combined pulsar and \gls{bbh} results, do place tighter restrictions on the parameters.  These arise due to a combination of \gls{ns} masses, which peak at $1.35\,M_{\odot}$ but incorporate a high mass tail, and \gls{ns} spins which exclude large spin values. The results exclude a binary with equal masses and give a \gls{bh} with a mass of $3.8^{+0.5}_{-0.6}\,M_{\odot}$ and a \gls{ns} with a mass of $1.4^{+0.2}_{-0.1}\,M_{\odot}$.  The spins are bounded closer to zero although anti-aligned spins remain possible, particularly for the \gls{bh}.

The Bayes factors between the different prior assumptions are not the focus of this work and the different models can be partially overlapping and not designed to be compared against each other. However the correction to the evidence from the analysis in \cite{gw230529} can be easily computed with the reweighting approach (see \cite{2019PhRvD.100l3017P} for a clear exposition) and we find no significant changes, with the highest effect coming from the astrophysical model with an increase in evidence of $\log_{10}\mathcal{B}\approx 0.5$, comparable to the other effects, such as choice of waveform model, mentioned in \cite{gw230529}.

\section{Discussion}
\label{sec:discussion}

This paper re-analyses the \gls{gw} event GW230529 with a range of astrophysically motivated priors on the masses and spins.  This work is complementary to that presented in \cite{gw230529}, where prior distributions derived from the handful of previously observed \gls{nsbh} observations were used.  Here, we have made use of priors derived from {population synthesis models of stellar binaries}, observations of pulsars in the galaxy and \gls{bbh} binaries through \gls{gw} observations.  

The first key-point is that the inferred masses and spins of the components of the binary depend critically upon the mass and spin priors used in the analysis and, in particular, the inferred mass distribution is highly dependent upon the spin prior.  The fact that the distributions are so reliant on the prior demonstrates that the uncertainties in the observations for this system are large, due to the relatively low \gls{snr} of the event.  Therefore, we cannot draw strong conclusions about the origin of this event.  

However, we also note that under three distinct sets of astrophysically well-motivated choices of mass and/or spin priors, we arrive at a similar conclusion: that the preferred progenitor of GW230529 was a binary composed of a \gls{ns} and a \gls{bh}, where the \gls{ns} is entirely consistent with the observed galactic population and the \gls{bh} lies at the upper end of the purported ``lower mass gap'' between $2.5$-$5 M_{\odot}$.

Even with the most optimistic (i.e. broadest) observationally motivated spin priors --- \gls{bbh} (non-zero spin distribution from gravitational waves catalogue GWTC-3) and the millisecond pulsar spins (not accounting for pulsar spin-down at merger), our results are unaltered.
We also conclude that while we cannot rule out any of the astrophysical models described in Section \ref{sec:modelA}, or say with certainty that the \gls{ns} was formed before the \gls{bh}, we can most definitely rule out the tidal spin-up of the \gls{bh}, hence rendering the lack of observed electromagnetic counterpart unsurprising \citep{Barbieri2020}. Higher signal-to-noise ratio, multiple-detector observation (for better sky-localization) and an order-of-magnitude closer events in future \gls{gw}  observing runs will provide improved ability to accurately measure the binary parameters, and increase the chance of observing electromagnetic counterparts to similar observations in the future.  The observation of low mass, high spin \glspl{bh} will provide evidence of tidal spin-up and also provide a greater chance of observing a counterpart. 

\section*{Acknowledgements}
We thank Vivek Venkatraman Krishnan, Mark Hannam and Alexandre G\"ottel for the useful discussions.
DC, FA, SF and VR are supported by the UK's Science and Technology Facilities Council grant ST/V005618/1. SAS is funded by UKRI Centre for Doctoral Training in Artificial Intelligence, Machine Learning \& Advanced Computing. The authors are grateful for computational resources provided by Cardiff University, which are supported by STFC grants ST/I006285/1 and ST/V005618/1. The astrophysical model grid was run in the OzSTAR high-performance supercomputer at  Swinburne  University of  Technology. 
OzSTAR is funded by Swinburne University of Technology and the National Collaborative Research Infrastructure Strategy (NCRIS). 
This research has made use of data or software obtained from the Gravitational Wave Open Science Center (gwosc.org), a service of the LIGO Scientific Collaboration, the Virgo Collaboration, and KAGRA. This material is based upon work supported by NSF's LIGO Laboratory which is a major facility fully funded by the National Science Foundation, as well as the Science and Technology Facilities Council (STFC) of the United Kingdom, the Max-Planck-Society (MPS), and the State of Niedersachsen/Germany for support of the construction of Advanced LIGO and construction and operation of the GEO600 detector. Additional support for Advanced LIGO was provided by the Australian Research Council. Virgo is funded, through the European Gravitational Observatory (EGO), the French Centre National de Recherche Scientifique (CNRS), the Italian Istituto Nazionale di Fisica Nucleare (INFN) and the Dutch Nikhef, with contributions by institutions from Belgium, Germany, Greece, Hungary, Ireland, Japan, Monaco, Poland, Portugal, Spain. KAGRA is supported by the Ministry of Education, Culture, Sports, Science and Technology (MEXT), Japan Society for the Promotion of Science (JSPS) in Japan; National Research Foundation (NRF) and the Ministry of Science and ICT (MSIT) in Korea; Academia Sinica (AS) and National Science and Technology Council (NSTC) in Taiwan.

\section*{Data Availability}

The data utilized for this work will be freely available upon reasonable request to the corresponding author.


\bibliographystyle{mnras}
\bibliography{papers}



\appendix


\bsp	
\label{lastpage}
\end{document}